\documentstyle[12pt,cite]{ioplppt}

\makeatletter

\@addtoreset{equation}{section}

\makeatother

\begin{document}

\jl{1}

%%%%%%%%%%%%%%%%%% TITLE %%%%%%%%%%%%%%
  \title{Separation of Variables\\
    in the BC-type Gaudin Magnet
    \ftnote{7}{Received on April 21, 1995}
    }[BC-type Gaudin Magnet]

%%%%%%%%%%%%%%%%%%%%%%% AUTHOR(S) %%%%%%%%%%%%%%%%%%%
  \author{Kazuhiro Hikami\ftnote{1}{
      hikami@monet.phys.s.u-tokyo.ac.jp
      }
    }

%%%%%%%%%%%%%%%%%%%%%%% ADDRESS %%%%%%%%%%%%%%%%%%%%
  \address{Department of Physics, Graduate School of Science,\\
    University of Tokyo,\\
    Hongo 7-3-1, Bunkyo, Tokyo 113, Japan.
    }

%%%%%%%%%%%%%%%%%%%%%% DATE %%%%%%%%%%%%%%%%%%%%%%%%%%%

%%%%%%%%%%%%%%%%%%%%%% ABSTRACT %%%%%%%%%%%%%%%%%%%%%%
\begin{abstract}
  The integrable system is introduced based on the Poisson $ rs $-matrix
  structure.
  This is a generalization of the Gaudin magnet,
  and in SL(2) case isomorphic
  to  the generalized  Neumann model.
  The separation of variables is  discussed for both classical and
  quantum case.
\end{abstract}

%%%%%%%%%%%%%%%%%%%%%% PACS %%%%%%%%%%%%%%%%%%%%%%%%%%

\maketitle
%%%%%%%%%%%%%%%%%%%%%%%%%%%%%%%%%%%%%%%%%%%%%%%%%%%%%%%

\section{Introduction}

The classical integrable systems has been formulated in terms of the
classical $ r $-matrix~\cite{FaTa87}.
In one sense, the system is proved to be integrable   when we can
perform the separation of variables;
the reduction of a
multi-dimensional system to a set of one-dimensional systems (see, for
review, Ref.~\cite{Sklyan95}).
Although the separation of variables has been widely
known as the
Hamiltonian-Jacobi equation,
Sklyanin proposed a new
technique (functional Bethe ansatz), which is closely related with
the (quantum) inverse scattering  method.
The functional Bethe ansatz method was first applied to classical
top~\cite{Skl85a}, and further applications, for the Toda lattice,
Gaudin magnet, and Heisenberg spin chain,
have been done.
This technique is a new tool in studying  integrable systems.

We briefly review the separation of variables for the Gaudin
magnet~\cite{Skl89} in classical case.
The fundamental formulation of   integrable system is based on
the classical $ r $-matrix
structure,
\begin{equation}
  \{ \stackrel{1}{{\rm {\bf L}}} \! (u) ,
  \stackrel{2}{{\rm {\bf L}}} \! (v) \}
  = [ {\rm {\bf r}}(u-v) ,
  \stackrel{1}{{\rm {\bf L}}} \! (u)
  +  \stackrel{2}{{\rm {\bf L}}} \! (v)
  ] .
  \label{llr}
\end{equation}
Here
we have used the standard notation,
$
  \stackrel{1}{{\rm {\bf L}}} \! \! (u) =
  {\rm {\bf L}}(u) \otimes 1
$
and
$
  \stackrel{2}{{\rm {\bf L}}} \! \! (v) = 1 \otimes {\rm {\bf L}}(v)
$.
As an example of the $ L $-matrix satisfying the linear Poisson
structure~(\ref{llr}), we can take
\begin{equation}
  {\rm {\bf L}}(u) =
  {\rm {\bf Z}}
  +
  \sum_{j=1}^N \frac{{\rm {\bf S}}_j}{u-z_j} .
  \label{original_l}
\end{equation}
Here $ {\rm {\bf S}}_j $ ($ j=1,\ldots,N $) is a classical
SL($n$) spin matrix, and its elements
$ S_j^{ab} $ ($a,b=1,\ldots,n$)
satisfy the Poisson relation,
\begin{equation}
  \{ S_j^{ab} , S_k^{cd} \}
  =
  \delta_{jk} \cdot
  \Bigl(
    \delta^{bc} \, S_j^{ad} - \delta^{da} \, S_j^{cb}
  \Bigr) ,
  \label{spin_poisson}
\end{equation}
and $ {\rm Tr} \,  {\rm {\bf S}}_j^2 $ is the Casimir element.
We suppose that matrix $ {\rm {\bf Z}} $ is  traceless,
$ {\rm Tr} \, {\rm {\bf Z}} = 0 $.
The $ L $-matrix~(\ref{original_l})
appears as a quasi-classical limit of the inhomogeneous Heisenberg
XXX-spin
chain~\cite{Jur89a,HKW92b}, and satisfies the linear Poisson
relation~(\ref{llr}) with the classical
$ r $-matrix,
\begin{equation}
  {\rm {\bf r}}(u) = \frac{{\rm {\bf P}}}{u} .
  \label{rational_r}
\end{equation}
Matrix $ {\rm {\bf P}} $ means  a permutation matrix,
$  P^{ab,cd} = \delta^{ad} \delta^{bc} $,
which satisfies
$ {\rm {\bf P}} \,  x \otimes y = y \otimes x $.
Note that $ r $-matrix~(\ref{rational_r}) is a rational solution of
the classical Yang-Baxter equation~\cite{BelaDrin81},
\begin{equation}
  [ {\rm {\bf r}}_{23}(v) , {\rm {\bf r}}_{12}(u) ]
  + [ {\rm {\bf r}}_{23}(v) , {\rm {\bf r}}_{13}(u+v) ]
  + [ {\rm {\bf r}}_{13}(u+v) , {\rm {\bf r}}_{12}(u) ] = 0 .
\end{equation}

Once the $ L $-matrix satisfies the Poisson structure~(\ref{llr}),
the model can be  proved to be integrable in the Liouville's sense.
If functions  of the $ L $-matrix,
$ \tau_m (u) $, are defined as
\begin{equation}
  \tau_m (u)  \equiv
  \frac{1}{m} \, {\rm Tr} \, {\rm {\bf L}}(u)^m ,
\end{equation}
one can see that $ \tau_m (u) $'s  are spectral invariant,
\begin{equation}
  \{ \tau_l (u) , \tau_m(v) \} = 0 .
  \label{tau_invo}
\end{equation}
The identity~(\ref{tau_invo}) shows that $ \tau_m (u) $'s are
generating functions  of the constants of motion.
The first non-trivial invariant follows from $ \tau_2(u) $;
\begin{equation}
  \tau_2(u) =
  \frac{1}{2} \, {\rm Tr} \, {\rm {\bf Z}}^2
  +
  \sum_{j=1}^N \frac{H_j}{u-z_j}
  +
  \frac{1}{2}
  \sum_{j=1}^N \frac{{\rm Tr} \, {\rm {\bf S}}_j {\rm {\bf S}}_j}
  {(u-z_j)^2} ,
\end{equation}
where $ H_j $ is the hamiltonian of the SL($n$) Gaudin magnet,
\begin{equation}
  H_j =
  {\rm Tr} \, {\rm {\bf Z}} {\rm {\bf S}}_j
  +
  \sum_{k \neq j}^N
  \frac{{\rm Tr} \, {\rm {\bf S}}_j {\rm {\bf S}}_k}{z_j-z_k} .
  \label{ham_gau}
\end{equation}
This model  was  introduced by Gaudin as an integrable spin system
with long-range interaction~\cite{Ga76}.
Due to the involutiveness of $ \tau_m (u) $, one sees that the
hamiltonian of the
SL($n$) Gaudin magnet is Poisson commutative,
\begin{equation}
  \{ H_j , H_k \} = 0 ,
  \qquad
  \mbox{for $ j,k=1,\ldots,N $.}
\end{equation}
The complete
integrability of the model in Liouville's sense can be proved
directly from~(\ref{tau_invo});
when we introduce quantities $ \tau_{m,j}^{\alpha} $ by
\begin{equation}
  \tau_m (u)
  = \frac{1}{m} \,  {\rm Tr} \, {\rm {\bf Z}}^m
  +
  \sum_{j=1}^N \sum_{\alpha=1}^m
  \frac{\tau_{m,j}^\alpha}{(u-z_j)^\alpha} ,
  \label{res_tau_m}
\end{equation}
we can see that the quantities $ \tau_{m,j}^\alpha $
($m=2,\ldots,N; j=1,\ldots,n; \alpha=1,\ldots,m-1$)
form a commutative family of $ N n(n-1)/2 $ independent hamiltonians.

For this type of the Gaudin magnet~(\ref{ham_gau}),
the separation of variable (functional Bethe ansatz)
has been  widely
studied
in both classical and quantum
cases~\cite{Skl89,BaTa92,Kuz92a,Kuz92c,AdamHarnHurt93,BrzeMacf94,KalKuzMil94}.
Let
$ {\cal A}({\rm {\bf L}}) $ and
$ {\cal B}({\rm {\bf L}}) $
be certain polynomials of  degree
$  n (n-1) /2 $ in  matrix elements $ L_{ab} $.
When we define
variables $ x_j $ and $ p_j $ by
\begin{equation}
  {\cal B}({\rm {\bf L}}(x_j)) = 0,
  \qquad
  p_j = {\cal A}({\rm {\bf L}}(x_j)),
\end{equation}
one sees that
variables $ x_j $ and $ p_j $
are canonically Poisson conjugate~\cite{Skl92,Scot94,Gekht95},
\begin{equation}
  \{ x_j , x_k \} = 0 ,
  \qquad
  \{ p_j , p_k \} = 0,
  \qquad
  \{ p_j , x_k \} = \delta_{jk} .
\end{equation}
This analysis, which is called the separation of variables, makes it
possible to  calculate the energy spectrum for the quantum Gaudin
magnet.

By this way, we can perform the separation of variables for integrable
systems formulated in the linear
Poisson relation~(\ref{llr}).
Some of  integrable systems,
e.g. nonlinear integrable equations  on finite segment,
are formulated in terms of  another
Poisson structure;
there exists `$rs$'-Poisson structure~\cite{Mai86,Skl87},
\begin{equation}
  \{ \stackrel{1}{{\rm {\bf L}}} \! (u) ,
  \stackrel{2}{{\rm {\bf L}}} \! (v) \}
  =
  [ {\rm {\bf r}}(u-v) ,
  \stackrel{1}{{\rm {\bf L}}} \! (u) +
  \stackrel{2}{{\rm {\bf L}}} \! (v) ]
  + [ {\rm {\bf s}}(u+v) ,
  - \! \stackrel{1}{{\rm {\bf L}}} \! (u) +
  \stackrel{2}{{\rm {\bf L}}} \! (v) ] .
  \label{reflection}
\end{equation}
Here $ {\rm {\bf r}} $ and $ {\rm {\bf s}} $ are  matrix structure
constants.
Remark that $ s $-matrix  depends on sum of spectral
parameters while
$ r $-matrix on difference.
%but  symmetries of the $ r $- and
%$ s $-matrices are different,
%\begin{eqnarray}
%  & & {\rm {\bf P}} \, {\rm {\bf r}}(u-v) \, {\rm {\bf P}} =
%  -{\rm {\bf r}}(v-u) , \\
%%
%  & & {\rm {\bf P}} \, {\rm {\bf s}}(u+v) \, {\rm {\bf P}} =
%  {\rm {\bf s}}(u+v) .
%\end{eqnarray}
The Poisson structure~(\ref{reflection}) can be viewed  as a
classical limit  of the boundary Yang-Baxter equation~\cite{Skl88},
which is used  to formulate the quantum spin chain
with open boundary.

In this paper  we shall study the separation of variables
for BC-type integrable system   formulated   by
the $rs$-Poisson structure~(\ref{reflection}).
In section~2, we introduce the BC-type $ {\rm SL}(n) $ Gaudin magnet.
We   give  classical  $ r $- and $ s $-matrices, and
prove  the integrability.
We relate the hamiltonian of the BC-type SL(2) Gaudin magnet with the
generalized Neumann model in section~3.
The separation of variables is also  studied.
In section~4,
we pay attention on quantum case.
The energy spectrum is given based on the  separation of variables.
Section~5 is devoted to conclusion and discussion.

%%%%%%%%%%%%%%%%%%%%%%%%%%%%%%%%%%%%%%
%\vspace{20mm}
\section{Gaudin Magnet with Boundary}
%\section{Description of the Model}

The quantum Gaudin magnet, whose hamiltonian has a
form~(\ref{ham_gau}),  was first introduced in Ref.~\cite{Ga76} as an
integrable
spin system with long-range interaction, and solved by use of the
coordinate Bethe ansatz.
As was reviewed in the previous section,
this original Gaudin magnet can be formulated with
the linear Poisson structure in classical case~(\ref{llr}).
In this section, we consider the $ {\rm SL}(n) $ Gaudin magnet with
boundary (BC-type Gaudin magnet).
The dynamical variables of this  model are
the classical SL($n$) spin
$ S_j^{ab} $ ($ a,b = 1, \ldots n $; $ j=1,\ldots,N $) satisfying the
Poisson bracket~(\ref{spin_poisson}).
Consider the modified $ L $-matrix,
\begin{equation}
  {\rm {\bf L}}(u) = \sum_{j=1}^N
  \biggl(
    \frac{{\rm {\bf S}}_j}{u-z_j}
    + \frac{\overline{{\rm {\bf S}}}_j}{u+z_j}
  \biggr) ,
  \label{lop}
\end{equation}
where   the ``reflected'' classical spin
$ \overline{{\rm {\bf S}}} $
is defined
as
\begin{equation}
  \overline{S}^{ab}
  = (-)^{a+b} S^{ab} .
\end{equation}
See the difference from the usual $ L $-operator~(\ref{original_l}).
The second term in (\ref{lop}) is due to the effect of the reflection;
the classical spin $ {\rm {\bf S}}_j $
is located at coordinate $ z_j $, while the ``reflected'' spin
$ \overline{{\rm {\bf S}}}_j $ is at $ -z_j $.
For this reason we say that the system has ``boundary''.
One can check easily that the modified $ L $-operator~(\ref{lop})
satisfies the linear
Poisson structure~(\ref{reflection}) with the rational
$ r $-matrix~(\ref{rational_r}) and $ s $-matrix,
\begin{equation}
  {\rm {\bf s}}(u)  =  \frac{\overline{{\rm {\bf P}}}}{u} ,
\end{equation}
where we use the notation,
$ \overline{P}^{ab,cd} = (-)^{a+b} \delta^{ad} \delta^{bc} $.
%The Poisson relation~(\ref{reflection}) can be rewritten in terms of
%matrix elements,
%\begin{eqnarray}
%  \{ L^{ab}(u) ,  L^{cd}(v) \}
%  & = &
%  \frac{1}{u-v}
%  \biggl(
%    \delta^{ad} \cdot
%    \Bigl( L^{cb}(u) - L^{cb}(v) \Bigr)
%    - \delta^{bc} \cdot
%    \Bigl( L^{ad}(u) - L^{ad}(v) \Bigr)
%  \biggr)
%  \nonumber \\
%  & & \mbox{ }
%  + \frac{1}{u+v}
%  \biggl(
%    (-)^{a+1} \delta^{ad} \cdot
%    \Bigl( (-)^c L^{cb}(u) + (-)^b L^{cb}(v) \Bigr)
%  \nonumber \\
%  & & \mbox{ } \; \; \;
%  +
%    (-)^b \delta^{bc} \cdot
%    \Bigl( (-)^d L^{ad}(u) + (-)^a L^{ad}(v) \Bigr)
%  \biggr) .
%\end{eqnarray}

Let us define  functions $ \tau_m(u) $ of the matrix
$ {\rm {\bf L}}(u) $~(\ref{lop}) as,
\begin{equation}
  \tau_m (u) = \frac{1}{m} \, {\rm Tr} \, {\rm {\bf L}}(u)^m .
  \label{tau_nu}
\end{equation}
{}From the Poisson structure~(\ref{reflection}),
it is  shown that
functions $ \tau_m (u) $
are the spectral invariants of dynamical system,
\begin{equation}
  \{ \tau_l (u) , \tau_m(v) \} = 0 .
  \label{tautau}
\end{equation}
The first nontrivial  invariant is given from $ \tau_2(u) $ as,
\begin{eqnarray}
  \tau_2(u)  & = &  \frac{1}{2} \,
  {\rm Tr} \, {\rm {\bf L}}(u)^2 \nonumber \\
  & = &  \sum_{j=1}^N
  \frac{2 z_j\, H_j}{(u-z_j)(u+z_j)}
  +
  \frac{1}{2} \, \sum_{j=1}^N
  \biggl(
    \frac{1}{(u-z_j)^2}
    + \frac{1}{(u+z_j)^2}
  \biggr)
  {\rm Tr} \, {\rm {\bf S}}_j^2
  ,
\end{eqnarray}
where  $ H_j $ has a form,
\begin{equation}
  H_j = \frac{{\rm Tr} \,
    ({\rm {\bf S}}_j \overline{{\rm {\bf S}}}_j)}{2 \, z_j}
  +  \sum_{k \neq j}^N
  \biggl(
    \frac{{\rm Tr} \, ({\rm {\bf S}}_j {\rm {\bf S}}_k)}{z_j-z_k}
    + \frac{{\rm Tr} \,
      ({\rm {\bf S}}_j \overline{{\rm {\bf S}}}_k)}{z_j+z_k}
  \biggr) .
  \label{bc_gaudin}
\end{equation}
The involutiveness of the spectral invariants $ \tau_m (u) $
indicates  the Poisson
commutativity  of the hamiltonian $ H_j $,
\begin{equation}
  \{ H_j , H_k \} = 0 ,
  \qquad
  \mbox{for $ j,k = 1,2,\ldots,N $.}
\end{equation}
We call $ H_j $ the hamiltonian of the SL($n$) Gaudin magnet with
boundary.
Different from the original Gaudin magnet~(\ref{ham_gau}),
the hamiltonian $ H_j $ includes interacting terms between
classical spins $ {\rm {\bf S}}_j $ and
``reflected'' spins $ \overline{{\rm {\bf S}}}_j $.
By
calculating the residues of $ \tau_m(u) $~(\ref{tau_nu})
as in the case of original Gaudin magnet~(\ref{res_tau_m}),
one can get ``higher-order'' hamiltonian $ \tau_{m,j}^\alpha $
of the Gaudin magnet.
The existence of commutative family $ \{ \tau_{m,j}^\alpha \} $
supports
the complete integrability of model in Liouville's sense.
%Note that we have another generating function of the conserved
%quantities,
%\begin{equation}
%  W(z,u)
%  =
%  {\rm det} \, (z- {\rm {\bf L}}(u) )
%  = \sum_{k=1}^n (-)^k t_k(u) z^{n-k} .
%\end{equation}

%%%%%%%%%%%%%%%%%%%%%%%%%%%%%%%%%%%%%%
%\vspace{20mm}
\section{Separation of Variables}

In this section we first  show the isomorphism of the $ N $-site SL(2)
Gaudin magnet and
the $ N $-dimensional generalized Neumann model, and then study the
separation of variables based on
the technique of Sklyanin.
In SL(2) case we can define the classical spin matrices
$ {\rm {\bf S}}_j $ and
$ \overline{{\rm {\bf S}}}_j $ in the
$ L $-operator~(\ref{lop}) as,
\begin{equation}
  {\rm {\bf S}}_j =
  \left(
    \begin{array}{cc}
      S_j^z & S_j^- \\
      S_j^+ & - S_j^z
    \end{array}
  \right) ,
  \qquad
  \overline{{\rm {\bf S}}}_j =
  \sigma^z {\rm {\bf S}}_j \sigma^z ,
\end{equation}
where $ \sigma^z $ is denoted as Pauli spin matrix.
These  spin variables satisfy the following Poisson relations;
\begin{equation}
  \{ S_j^z , S_k^\pm \} = \pm \delta_{jk} \, S_j^\pm ,
  \qquad
  \{ S_j^+ , S_k^- \} = 2 \delta_{jk} \, S_j^z .
\end{equation}
Above  Poisson structures for spin variables can be realized with
new variables
$ x_j $ and $ p_j $ as,
\begin{equation}
  S_j^+ = \frac{{\rm i}}{2} x_j^2 ,
  \qquad
  S_j^- = \frac{{\rm i}}{2} p_j^2 ,
  \qquad
  S_j^z = - \frac{1}{2} x_j p_j ,
\end{equation}
where  $ \{ x_j,  p_j | j = 1, \ldots, N \} $
are canonical variables  satisfying the Poisson
relations;
$ \{ x_j , x_k \} = \{ p_j , p_k \} = 0 $,
and
$ \{ x_j , p_k \} = \delta_{jk} $.
In this case the Casimir element is set to be zero,
$ {\rm Tr} \, {\rm {\bf S}}_j^2 = 0 $.
In terms of   canonical variables, one can obtain the hamiltonian
from the spectral invariant
$ \tau_2(u) $,
\begin{eqnarray}
  \tau_2(u) & = & \frac{1}{2} \, {\rm Tr} \, {\rm {\bf L}}(u)^2
  \nonumber \\
  & = & \sum_{j=1}^N \frac{z_j}{(u-z_j)(u+z_j)} H_j ,
\end{eqnarray}
where $ H_j$ is  calculated  as,
\begin{equation}
  H_j =
  \frac{x_j^2 p_j^2}{z_j}
  -
  \frac{1}{2}
  \sum_{k \neq j}^N
  \biggl(
    \frac{(p_j x_k - x_j p_k)^2}{z_j-z_k}
    - \frac{(p_j x_k + x_j p_k)^2}{z_j+z_k}
  \biggr) .
  \label{neumann}
\end{equation}
This hamiltonian can be viewed as  generalization of the Neumann
model~\cite{BaTa92,Kuz92c}.
This proves the fact  that the $ N $-dimensional generalized Neumann
model~(\ref{neumann}) is isomorphic to the $ N $-site BC-type SL(2)
Gaudin magnet~(\ref{bc_gaudin}).

Now we study the separation of variables for the generalized Neumann
model~(\ref{neumann}).
Define $ A(u) $ and $ B(u) $ as
\begin{equation}
  A(u) =  L_{11}(u) ,
  \qquad
  B(u) =  L_{12}(u) .
\end{equation}
The linear Poisson structure~(\ref{reflection}) includes
relations between   functions
$ A(u) $ and $ B(u) $;
\begin{equation}
  \label{aabb}
  \begin{array}{l}
    \{ A(u) , A(v) \} = 0 ,  \qquad
    \{ B(u) , B(v) \} = 0 ,  \qquad \\
    \noalign{\vskip 3mm}
    \displaystyle{
      \{ A(u) , B(v) \}
      =   \frac{2 \, u}{(u-v)(u+v)}
      \Bigl(
        B(u) - B(v)
      \Bigr) .
      }
  \end{array}
\end{equation}
Both matrix elements, $ A(u) $ and $ B(u) $,  Poisson-commute among
themselves.

We choose separable coordinates as zeros of the off-diagonal element
of $ L $-operator,
\begin{equation}
  B(\pm u_\alpha) = 0 ,
  \qquad
  \mbox{for $ \alpha=1,2,\ldots,N-1$.}
  \label{bzero}
\end{equation}
Note that $ u=-u_\alpha $ is  solution if
$ u=u_\alpha $ solves equation
$ B(u) = 0 $.
By use of  a   set of variables $ u_\alpha $,
we further introduce  the canonical variables by
\begin{equation}
  v_\alpha \equiv A(u_\alpha) .
\end{equation}
{}From the Poisson relations~(\ref{aabb})
one  can see that variables $ u_\alpha $ and $ v_\alpha $ are
canonically conjugate,
\begin{equation}
  \{ u_\alpha , u_\beta \}  =  0 , \qquad
  \{ v_\alpha , v_\beta \}  = 0 , \qquad
  \{ u_\alpha , v_\beta \}  =  \delta_{\alpha\beta} .
  \label{uv_relation}
\end{equation}
The first two Poisson relations  are easy to be proved.
The third relation   follows from
\[
  \{ u_\alpha, v_\beta \}
  = \lim_{u \to u_\alpha}
  \frac{1}{B^{'}(u)}
  \biggl(
    \frac{1}{u_\beta-u} + \frac{1}{u_\beta+u}
  \biggr) \, B(u) .
\]
The Poisson relations~(\ref{uv_relation}) show that $ u_\alpha $
and $ v_\alpha $ are canonically conjugate variables, and that the
generalized Neumann model~(\ref{neumann}) is separated by transforming
the dynamical variables as
\[
    \{ x_j, p_j | j=1,\ldots,N \}
  \to
  \{ u_\alpha, v_\alpha | j=1,\ldots,N-1\} .
%  \;
%  z \equiv \sum_j z_j p_j^2 , \;
%  \overline{z} \equiv \sum_j \frac{1}{2} (z_j p_j)^{-1} x_j .
\]
With variables $ u_\alpha $ and $ v_\alpha $,
the action $ W $ of the generalized Neumann model~(\ref{neumann}) is
written  as separated form,
\begin{equation}
  W = \sum_{\alpha=1}^{N-1}
  \int v_\alpha {\rm d}u_\alpha .
\end{equation}
The separated variables $ u_\alpha $ and $ v_\alpha $
can be written
explicitly in terms of $ x_j $ and $ p_j $.
By definition (\ref{bzero}) we can solve the coordinates $ p_j $ as
\begin{equation}
  p_j^2  =  \frac{z}{z_j} \cdot
  \frac{
    \displaystyle{
      \prod_{\alpha=1}^{N-1}(z_j-u_\alpha)(z_j+u_\alpha)
      }
    }
  {\displaystyle{
      \prod_{k \neq j}^N (z_j-z_k)(z_j+z_k)
      }
    }
  ,
\end{equation}
with
$ z \equiv \sum_j z_j p_j^2 $.
The canonically conjugate variables  $ v_\alpha $ are also solved as
\begin{equation}
  v_\alpha  \equiv A(u_\alpha)
  =
  -
  \sum_{j=1}^N
    \frac{u_\alpha \, x_j \, p_j}
    {(u_\alpha-z_j)(u_\alpha+z_j)} .
\end{equation}
Remark that
the relation between
$ \{ u_\alpha, v_\alpha \} $ and the hamiltonian $ H_j $
is given  from the spectral invariants
$ \tau_2(u = u_\alpha) $,
\begin{equation}
  v_\alpha^2
  =
  \sum_{j=1}^N
  \frac{z_j H_j}{(u_\alpha-z_j)(u_\alpha+z_j)} .
  \label{cl_vh}
\end{equation}

%%%%%%%%%%%%%%%%%%%%%%%%%%%%%%%%%%%%%%
%\vspace{20mm}
\section{Quantum Case}

In this section we consider to quantize the BC-type Gaudin
magnet~(\ref{bc_gaudin}).
For brevity we study SL(2) case.
We set the $ L $-operator as
\begin{equation}
  {\rm {\bf L}}(u)
  =
%  {\rm {\bf K}}
%  +
  \sum_{j=1}^N
  \biggl(
    \frac{{\rm {\bf S}}_j}{u-z_j}
    + \frac{\overline{\rm {\bf S}}_j}{u+z_j}
  \biggr) ,
  \label{quantum_l}
\end{equation}
where the spin operator $ {\rm {\bf S}}_j $
and reflected spin operator $ \overline{{\rm {\bf S}}}_j $
are defined as
\begin{equation}
  {\rm {\bf S}}_j
  =
  \left(
    \begin{array}{cc}
      S^z_j &   S^-_j \\
      S^+_j & - S^z_j
    \end{array}
  \right),
  \qquad
  \overline{{\rm {\bf S}}}_j
  =
  \sigma^z {\rm {\bf S}}_j \sigma^z
  =
  \left(
    \begin{array}{cc}
        S^z_j & - S^-_j \\
      - S^+_j & - S^z_j
    \end{array}
  \right) .
\end{equation}
Here operators $ S^z_j $ and $ S^\pm_j $ denote bases of the su(2) Lie
algebra,
and they  satisfy the commutation relations,
\begin{equation}
  \begin{array}{l}
    [ S_j^z , S_k^\pm ] = \pm S_j^\pm \delta_{jk},
    \qquad
    [ S_j^+ , S_k^- ] = 2 S_j^z \delta_{jk} , \\
    \noalign{\vskip 3mm}
    \displaystyle{
      ( S_j^z )^2 + \frac{1}{2}( S_j^+ S_j^- + S_j^- S_j^+ )
      = \ell_j (\ell_j+1) ,
      }
    \qquad
    \ell_j \in {\rm {\bf Z}}_+/2 .
  \end{array}
\end{equation}
We have set  $ \ell_j $ as  spin of the $ j $-th site.
%In~(\ref{quantum_l}) we also introduce the constant matrix
%$ {\rm {\bf K}} $,
%\begin{equation}
%  {\rm {\bf K}} =
%  \left(
%    \begin{array}{cc}
%      & \xi^+ \\
%      \xi^- &
%    \end{array}
%  \right) .
%\end{equation}
One can check
from direct calculations
that the $ L $-operator~(\ref{quantum_l}) satisfies the
quantum analogue of the linear Poisson structure~(\ref{reflection}),
\begin{equation}
  [ \stackrel{1}{{\rm {\bf L}}} \! (u) ,
  \stackrel{2}{{\rm {\bf L}}} \! (v) ]
  =
  [ {\rm {\bf r}}(u-v) , \stackrel{1}{{\rm {\bf L}}} \! (u)
  + \stackrel{2}{{\rm {\bf L}}} \! (v) ]
  +
  [ {\rm {\bf s}}(u+v) , - \! \stackrel{1}{{\rm {\bf L}}} \! (u)
  + \stackrel{2}{{\rm {\bf L}}} \! (v) ] .
  \label{quantum_rs}
\end{equation}
In this case the constant $ r $- and $ s $-matrices
are defined as
\begin{equation}
  {\rm {\bf r}}(u) = - \frac{{\rm {\bf P}}}{u} ,
  \qquad
  {\rm {\bf s}}(u) = \,
  \stackrel{1}{\sigma^z}  {\rm {\bf r}}(u)
  \stackrel{1}{\sigma^z} .
\end{equation}

The conserved operators are generated by the same way in the classical
case; trace of the $ L $-operator~(\ref{tau_nu}).
We  get  the first non-trivial operator from
$ \hat{\tau}_2(u) $,
\begin{eqnarray}
  \hat{\tau}_2(u)
  & = & \frac{1}{2} \, {\rm Tr} \, {\rm {\bf L}}(u)^2 \nonumber \\
  & = &
%  \frac{1}{2} \, {\rm Tr} \, {\rm {\bf K}}^2
%  +
  \sum_{j=1}^N
  \frac{2 \, z_j}{(u-z_j)(u+z_j)} \hat{H}_j
  + \sum_{j=1}^N
  \frac{4 \, z_j^2 \,  \ell_j \, (\ell_j+1)}{(u-z_j)^2 \, (u+z_j)^2} .
  \label{q_tau2}
\end{eqnarray}
Here the quantum operator $ \hat{H}_j $ is the hamiltonian of the
quantum
BC-type Gaudin magnet,
\begin{equation}
  \hat{H}_j =
%  {\rm Tr} \, {\rm {\bf K}} {\rm {\bf S}}_j
%  +
  \frac{(S_j^z)^2}{2 \, z_j}
  + \sum_{k \neq j}^N
  \biggl(
    \frac{{\rm Tr}\, {\rm {\bf S}}_j {\rm {\bf S}}_k}{z_j-z_k}
    + \frac{{\rm Tr} \, {\rm {\bf S}}_j
      \overline{{\rm {\bf S}}}_k}{z_j+z_k}
  \biggr)  .
  \label{ham_bc_gaudin}
\end{equation}
{}From  the commutativity of the generating function
$ \hat{\tau}_2(u) $, we can see that the operators $ \hat{H}_j $ are
commutative,
\begin{equation}
  [ \hat{H}_j , \hat{H}_k ] = 0 ,
  \qquad
  \mbox{for $j,k=1,\ldots,N,$}
\end{equation}
which proves the quantum integrability of the BC-type Gaudin magnet.
Note that the operator $ \hat{H}_j $~(\ref{ham_bc_gaudin})
has been appeared in
recent studies of the generalized Knizhnik-Zamolodchikov (KZ)
equation~\cite{Chere92b,Hikam95a}.

The separation of variables for quantum case
can be performed as follows.
When we define operator $ A(u) $ and $ B(u) $ as
\[
  A(u) = L_{11}(u) ,
  \qquad
  B(u) = L_{12}(u) ,
\]
we  obtain from the quantum $rs$-structure~(\ref{quantum_rs})
that
the commutation relations among operators
$ A(u) $ and $ B(u) $
can be written  as,
\begin{equation}
  \begin{array}{l}
    [ A(u) , A(v) ] = 0,
    \qquad
    [ B(u) , B(v) ] = 0,
    \\
    \noalign{\vskip 3mm}
    \displaystyle{
      [ A(u) , B(v) ]
      = \frac{2 \, u}{(u-v)(u+v)}
      \Bigl(
        B(u) - B(v)
      \Bigr) .
      }
  \end{array}
\end{equation}
All the calculation is essentially same with the classical case.
In the quantum case, we can also introduce the ``canonical operators''
$  u_\alpha $ and $ v_\alpha $ by,
\begin{equation}
  \begin{array}{l}
    B(\pm u_\alpha) = 0 ,
%    \qquad
%    \mbox{for $\alpha=1,\ldots,N$,}
    \\
    \noalign{\vskip 3mm}
    v_\alpha = A(u_\alpha) .
  \end{array}
  \label{sepa_uuvv}
\end{equation}
These operators $ u_\alpha $ and $ v_\alpha $ satisfy the commutation
relations,
\begin{equation}
    [u_\alpha, u_\beta ] = 0 ,
    \qquad
    [v_\alpha , v_\beta ] = 0 ,
%
%    \noalign{\vskip 3mm}
%
    \qquad
    [u_\alpha , v_\beta ] = \delta_{\alpha \beta} .
    \label{uuvv}
\end{equation}

To perform the separation of variables~(\ref{sepa_uuvv})
for the quantum BC-type Gaudin magnet,
we use the realization
of the spin operators, $ S_j^z $ and $ S_j^\pm $, as
\begin{equation}
  S_j^z = - x_j \frac{\partial}{\partial x_j} + \ell_j ,
  \qquad
  S_j^- = x_j ,
  \qquad
  S_j^+ = - x_j \frac{\partial^2}{\partial x_j^2} + 2 \ell_j
  \frac{\partial}{\partial x_j} .
  \label{spin_real}
\end{equation}
With  this realization, the functional equation,
$ B(\pm u_\alpha) = 0 $,
does not include differential operator,  and
can be solved easily.
The result is
%$
%  \{ x_j \}
%  \to
%  \{ u_\alpha \}
%$.
%By using the definition of $ u_\alpha $, we have
\begin{equation}
  x_j =
  \frac{z}{2 z_j}
  \frac{
    \displaystyle{
      \prod_{\alpha=1}^{N-1} (z_j-u_\alpha)(z_j+u_\alpha)
      }
    }
  {
    \displaystyle{
      \prod_{k\neq j}^N (z_j-z_k)(z_j+z_k)
      }
    },
\end{equation}
where we  set
$ z \equiv \sum_k  z_k x_k $.
%When we use the above identity and the spin realization
%in~(\ref{spin_real}),
When we change the variables from $ \{ x_j \} $ to
$ \{ u_\alpha \} $,
we can see
that  operator $ v_\alpha $ is represented in terms of
$ u_\alpha $ as
\begin{equation}
  v_\alpha = - \frac{\partial}{\partial u_\alpha}
  + \Lambda(u_\alpha) ,
  \label{v_alpha}
\end{equation}
where we use  function $ \Lambda(u) $ as
\begin{equation}
  \Lambda(u) = \sum_{j=1}^N
  \frac{ 2 \, u \, \ell_j}{(u-z_j)(u+z_j)} .
\end{equation}
Remark that identification of operator
$ v_\alpha $~(\ref{v_alpha})
is consistent with  the commutation
relations~(\ref{uuvv}).
A set of operators $ \{ u_\alpha , v_\alpha \} $ is called the
separated operator.

We have completed in  separating  variables for quantum SL(2) Gaudin
magnet.
In the rest of this section,
we show that the energy spectrum for the Gaudin magnet can be
calculated
from the functional Bethe ansatz.
By definition of the generating function,
$ \hat{\tau}_2 (u) = \frac{1}{2} \,
{\rm Tr} \, {\rm {\bf L}}(u)^2 $,
one can see that
\begin{equation}
  v_\alpha^2 - \hat{\tau}_2(u_\alpha) = 0 ,
  \label{tau2_op}
\end{equation}
which corresponds, in classical case, to~(\ref{cl_vh}).
With the operator
realization of $ v_\alpha $ obtained in~(\ref{v_alpha}),
we can read off the identity~(\ref{tau2_op}) as
the   differential operator  for   the
separated spectral problem,
\begin{equation}
  \psi^{\prime \prime}(u)
  - 2 \Lambda(u) \psi^{\prime}(u)
  + \Bigl( \Lambda^2(u) - \Lambda^{\prime}(u) \Bigr) \psi(u)
  = \tau_2(u) \psi(u) ,
  \label{g_lame}
\end{equation}
where $ \tau_2(u) $ is an eigenvalue of
operator $ \hat{\tau}_2(u) $~(\ref{q_tau2}).
This equation
can be seen as  a generalized Lam{\'e} equation~\cite{Kuz92a}.
To solve this  second order differential equation~(\ref{g_lame}),
we  assume that the wave function $ \psi(u) $ is a polynomial of
$ u $, and that zeros of $ \psi(u) $ are  denoted as
$ \pm \lambda_\alpha $~\cite{Skl89},
\begin{equation}
  \psi(u) = \prod_{\alpha=1}^M (u-\lambda_\alpha)(u+\lambda_\alpha) .
\end{equation}
Substituting the wave function $ \psi(u) $ into differential
equation~(\ref{g_lame}),
we can see that the eigenvalue $ E_j $ of $ \hat{H}_j $ is given by,
\begin{equation}
  E_j = - 2 \, \chi(z_j) \, \ell_j
  - \frac{\ell_j}{z_j}
  + \sum_{k \neq j}^N \frac{4 \, z_j \, \ell_j \,
    \ell_k}{(z_j-z_k) (z_j+z_k)} ,
\end{equation}
where function $ \chi(u) $ is defined from the wave function
$ \psi(u) $ as
\begin{equation}
  \chi(u) =
  \frac{{\rm d}}{{\rm d}u} \log \psi(u)
  = \sum_{\alpha=1}^M \frac{ 2 \, u
    }{(u-\lambda_\alpha)(u+\lambda_\alpha)} .
\end{equation}
Notice that the zeros of the wave function $ \psi(u) $
%$ \lambda_\alpha $,
should be  fixed to satisfy a
set of equations,
\begin{equation}
  \Lambda(\lambda_\alpha)
  = \sum_{\beta \neq \alpha}^M \frac{2 \, \lambda_\alpha}
  {(\lambda_\alpha - \lambda_\beta) (\lambda_\alpha + \lambda_\beta)}
  + \frac{1}{2 \, \lambda_\alpha} ,
  \qquad
  \mbox{for $\alpha=1,\ldots,M$,}
\end{equation}
which
follows from the conditions in cancellation of
the residues  at
$ u=\lambda_\alpha $ in~(\ref{g_lame}).
This equation is a quasi-classical limit of the Bethe ansatz equation
for open boundary spin chain, and plays a crucial role in construction
of the integral solution for the generalized KZ
equation~\cite{Hikam95a}.

%%%%%%%%%%%%%%%%%%%%%%%%%%%%%%%%%%%%%%
%\vspace{20mm}
\section{Discussion}

%We only considered the separation of variables for SL(2) type.
%Recent studies in Refs.~\cite{Scot94,Gekht95} establish the separation
%of variables for SL($n$) type.

We have introduced the generalized Gaudin magnet in this paper.
The hamiltonian is written as
\[
  H_j = \frac{{\rm Tr} \,
    ({\rm {\bf S}}_j \overline{{\rm {\bf S}}}_j)}{2 \, z_j}
  +  \sum_{k \neq j}^N
  \biggl(
    \frac{{\rm Tr} \, ({\rm {\bf S}}_j {\rm {\bf S}}_k)}{z_j-z_k}
    + \frac{{\rm Tr} \,
      ({\rm {\bf S}}_j \overline{{\rm {\bf S}}}_k)}{z_j+z_k}
  \biggr) ,
\]
This  model
can be regarded as the Gaudin magnet  with boundary, and  be
formulated in terms of ``classical'' reflection
equation~(\ref{reflection}).
%Note that the BC-type Gaudin magnet is closely connected with the
%generalized Knizhnik-Zamolodchikov equation.
As in the case of the original Gaudin magnet, this model has many
interesting aspects;
in particular for  SL(2) case,  the model is proved to
be isomorphic to the generalized Neumann model.
In both classical and quantum cases we have performed the separation
of variables.
In this analysis the eigenvalue problem can be reduced to the
second-order differential equation (Lam\'{e}  equation).
We can obtain the so-called
quasi-classical  Bethe ansatz equation from this
differential equation.
The XXZ-Gaudin magnet with boundary will be analysed in the same
method.

The point of the separation of variables (functional Bethe ansatz) is
to take  zeros of the wave-function $ \psi(u) $;
the zeros may be identified with the ``rapidities'' of the spin-wave
from the view point of the inverse scattering method.
This kind of analysis   was used   in recent studies
of the Asbel-Hofstadter problem~\cite{WiegZabr94b}.
{}From the viewpoint of the $ q $-polynomial theory
the
Askey-Wilson polynomial may be related with quantum
$ rs $-structure~\cite{WiegZabr95a}.

%%%%%%%%%%%%%%%%%%%%%%%%%%%%%%%%%%%%%%

\ack

The author would like to thank Miki~Wadati for kind interests in this
work.
He also thanks E.~K.~Sklyanin for stimulating and useful discussions.

%%%%%%%%%%%%%%%%%%%%%%%%%%%%%%%%%%%%%%

\end{document}